\documentclass[twocolumn,showpacs,superscriptaddress,amsmath,amssymb,prl]{revtex4}

\usepackage{graphicx}%
\usepackage{dcolumn}%

\begin{document}

\title{Constituents of the \textquotedblleft kink\textquotedblright~in high-$T_c$ cuprates}

\author{A. A. Kordyuk}
\affiliation{Leibniz-Institut f\"ur Festk\"orper- und Werkstoffforschung Dresden, 01069 Dresden, Germany}
\affiliation{Institute of Metal Physics of National Academy of Sciences of Ukraine, 03142 Kyiv, Ukraine}

\author{S. V. Borisenko}
\author{V. B. Zabolotnyy}
\author{J. Geck}
\author{M. Knupfer}
\author{J. Fink}
\author{B. B\"uchner}
\affiliation{Leibniz-Institut f\"ur Festk\"orper- und Werkstoffforschung Dresden, 01069 Dresden, Germany}

\author{C. T. Lin}
\author{B. Keimer}
\affiliation{Max-Planck Institut f\"ur Festk\"orperforschung, 70569 Stuttgart, Germany}

\author{H. Berger}
\affiliation{Institut de Physique de la Mati\'ere Complexe, EPFL, 1015 Lausanne, Switzerland}

\author{Seiki Komiya}
\author{Yoichi Ando}
\affiliation{Central Research Institute of Electric Power Industry, Komae, Tokyo 201-8511, Japan}

\date{October 1, 2005}%

\begin{abstract}
Applying the Kramers-Kronig consistent procedure, developed earlier, we investigate in details the formation of the quasiparticle spectrum along the nodal direction of high-$T_c$ cuprates. The heavily discussed \textquotedblleft 70 meV kink\textquotedblright~on the renormalized dispersion exhibits a strong temperature and doping dependence when purified from structural effects. This dependence is well understood in terms of fermionic and bosonic constituents of the self-energy. The latter follows the evolution of the spin-fluctuation spectrum, emerging below $T$* and sharpening below $T_c$, and is the main responsible for the formation of the kink in question.  
\end{abstract}


\preprint{\textit{xxx}}

\maketitle

The nodal direction is thought as a simplest place in the Brillouin zone of the high-$T_c$ cuprates where the electron renormalization effects can be most easily understood. However, since the discovery of an energy scale in the experimental dispersion \cite{VallaSci1999, BogdanovPRL2000, KaminskiPRL2001}, a so-called \textquotedblleft 70 meV kink", its origin remains a matter of extensive debates \cite{EschrigPRL2000, JohnsonPRL01, LanzaraNature01, ZhouNature03, ZhouPRL2005, KoitzschPRB2004, KordyukPRL2004, KordyukPRB2004, KordyukPRB2005, Odd, Zn}, which now have mainly converged into a vital dilemma: phonons \textit{vs.}~spin-fluctuations \cite{DamascelliRMP03}. Historically, the kink has been associated with a coupling to the magnetic resonance mode because of its energy and doping dependence \cite{BogdanovPRL2000}, its seemingly smooth evolution into a spectral dip when moving to the antinodal region \cite{KaminskiPRL2001, EschrigPRL2000}, and its temperature dependence (emerging below $T_c$) \cite{JohnsonPRL01}. At the same time, the persistence of the effect above $T_c$ reported by another group \cite{BogdanovPRL2000} was taken as an argument against the resonance mode scenario. Moreover, a visual \textquotedblleft ubiquity" of the kink for a number of families of cuprates in a wide range of doping and temperature \cite{LanzaraNature01} and recently found similarity between a fine structure seen in dispersion to an expected phonon spectrum \cite{ZhouPRL2005} have made a strong claim in favor of phonon scenario. However, also recently, we have reported a careful investigation of the scattering rate kink \cite{KordyukPRL2004}, which is a simple consequence of the Kramers-Kronig (KK) relation between the real and imaginary parts of the electron self-energy \cite{KordyukPRB2005}, and which has appeared to be strongly doping and temperature ($xT$) dependent and, therefore, questions the phonon scenario. Moreover, the odd parity \cite{Odd} and strong dependence on Zn impurities \cite{Zn} of the nodal scattering form solid arguments for the magnetic scenario. Thus, from a number of arguments from both sides, it seems that the studies of the nature of the nodal kink have brought us to a stalemate, and an evident way to resolve it is to turn from a qualitative consideration of the kink effect to its quantitative analysis to derive the parameters of the bosonic spectrum that will allow to unambiguously identify its origin.


Recently we have developed a KK-consistent procedure \cite{KordyukPRB2005} which allows to extract both the real and imaginary parts of the self-energy, as well as the underlying bare dispersion from the photoemission data, and, thus, to place the kink problem into a quantitative domain. Subsequently, we have applied this self-consistent procedure to a number of nodal photoemission spectra measured at different temperatures and doping levels. Here we present the result of this investigation. We give a quantitative summary on the evolution of the nodal quasiparticle self-energy with doping and temperature and conclude about a determinative role of the spin-fluctuations in this evolution.

\begin{figure}[!t]
\includegraphics[width=8.6cm]{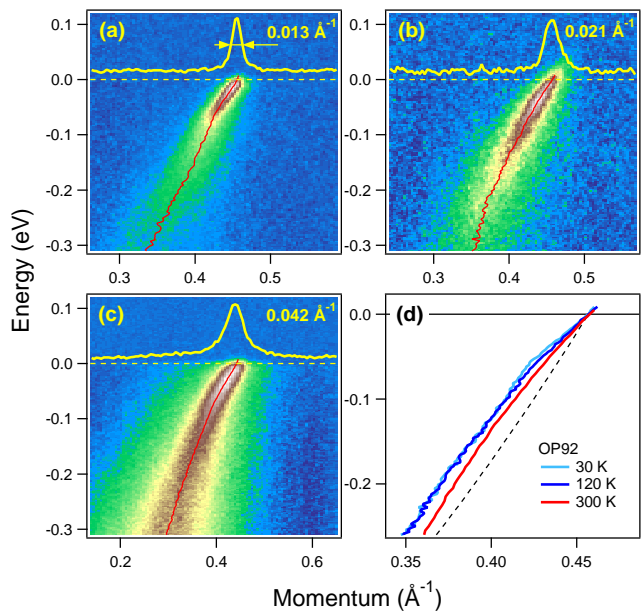}%
\caption{\label{Images} Photoemission spectra for an optimally doped Bi-2212 measured at 30 K (a), and 120 K (b), and an overdoped LSCO measured at 40 K (c); some examples of experimental dispersions for OP Bi-2212 to illustrate a vanish of the kink with rising the temperature, dashed line represents the bare dispersion.}
\end{figure}

We have analyzed the spectra from Bi$_2$Sr$_2$CaCu$_2$O$_{8+\delta}$: pure Bi-2212 and superstructure free Bi(Pb)-2212, and La$_{2-x}$Sr$_x$CuO$_4$ (LSCO) samples (we mark the samples according their doping and $T_c$). Some examples are shown in Fig.~\ref{Images} (a)-(c). The self-consistency requirement sets rigorous constrains on quality of the experimental data \cite{KordyukPRB2005}. The widths of the $E_F$ momentum distribution curves (MDC) are shown to illustrate the quality of the data we analyze. Under \textquotedblleft quality" we imply here not only good experimental statistics and overall resolution but also a purity of spectra from artificial components, e.g., due to superstructure or sample inhomogeneities. A special complication comes from the bilayer splitting in Bi-2212, non-vanishing along the nodal direction \cite{KordyukPRB2004}, that practically means that only spectra measured with 27 eV photons in the 1st Brillouin zone, when the photoemission from the bonding band is highly suppressed, can pass the KK-criterion \cite{KordyukPRB2005}. Therefore, all the presented spectra of Bi-2212 have been measured with 27 eV photons.

Fig.~\ref{Images} (d) is intended to illustrate a disappearance of the kink with rising the temperature when the bonding band is suppressed by a proper choice of the excitation energy $h\nu =$ 27 eV: at 300 K only a hump on the dispersion remains. Comparing two \textquotedblleft kinked" dispersions, it is clear that in order to get a reliable information about evolution of the interactions which form the kink with temperature and doping, one should consider the difference of the data derived functions, such as experimental dispersions, that requires an exceptional experimental statistics. All the mentioned constrains drastically decrease the amount of experimental data suitable for the precise analysis. Though a positive thing is that the bare dispersion is determined for each particular cut in the Brillouin zone and as far as this cut goes through the nodal point, the extracted self-energy is not very sensitive to its angular deviations from $(0,0)$-$(\pi,\pi)$ direction.

Using the aforementioned procedure, thoroughly described in \cite{KordyukPRB2005}, the bare electron dispersion $\varepsilon(k)$ can be extracted from the photoemission intensity distributions similar to those shown in Fig.~\ref{Images} (a)-(c). Consequently, both parts of the self-energy, $\Sigma'(\omega) = \omega - \varepsilon(k_m)$, and $\Sigma''(\omega) = [\varepsilon(k_1) - \varepsilon(k_2)]/2 \approx -v_F W$, are accessible as well. Here, for each $\omega$, three momentums, $k_m(\omega)$, $k_1(\omega)$, and $k_2(\omega)$, are defined for the MDC $A(k)$ as $A(k_m) = \max[A(k)]$, $A(k_{1,2}) = \max[A(k)]/2$; the MDC width $2W = k_2 - k_1$; $v_F$ is the bare Fermi velocity. In the following we focus on the self-energy functions derived from photoemission data.

\begin{figure*}[t]
\includegraphics[width=5.5cm]{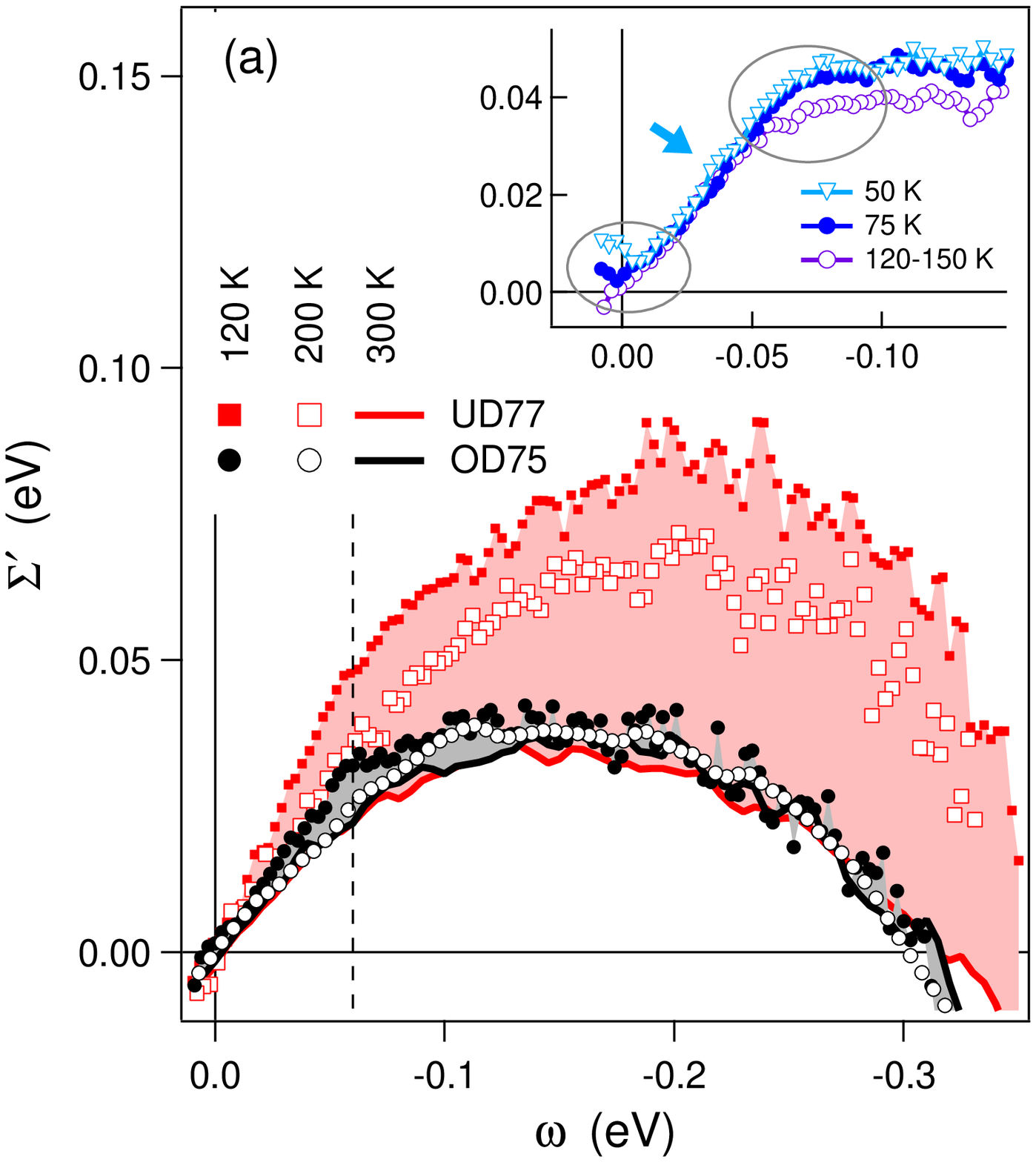}%
\includegraphics[width=4.95cm]{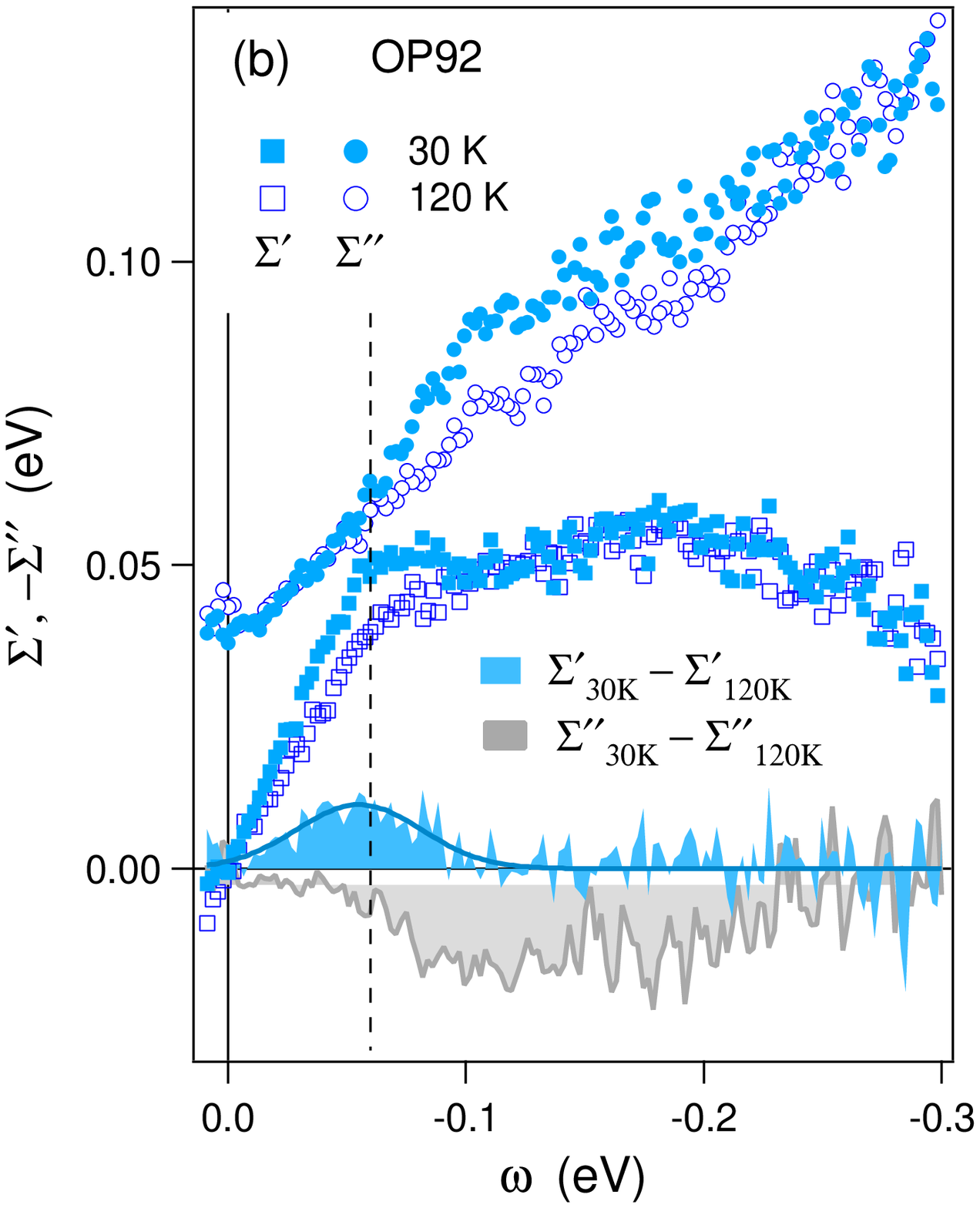}%
\includegraphics[width=7.65cm]{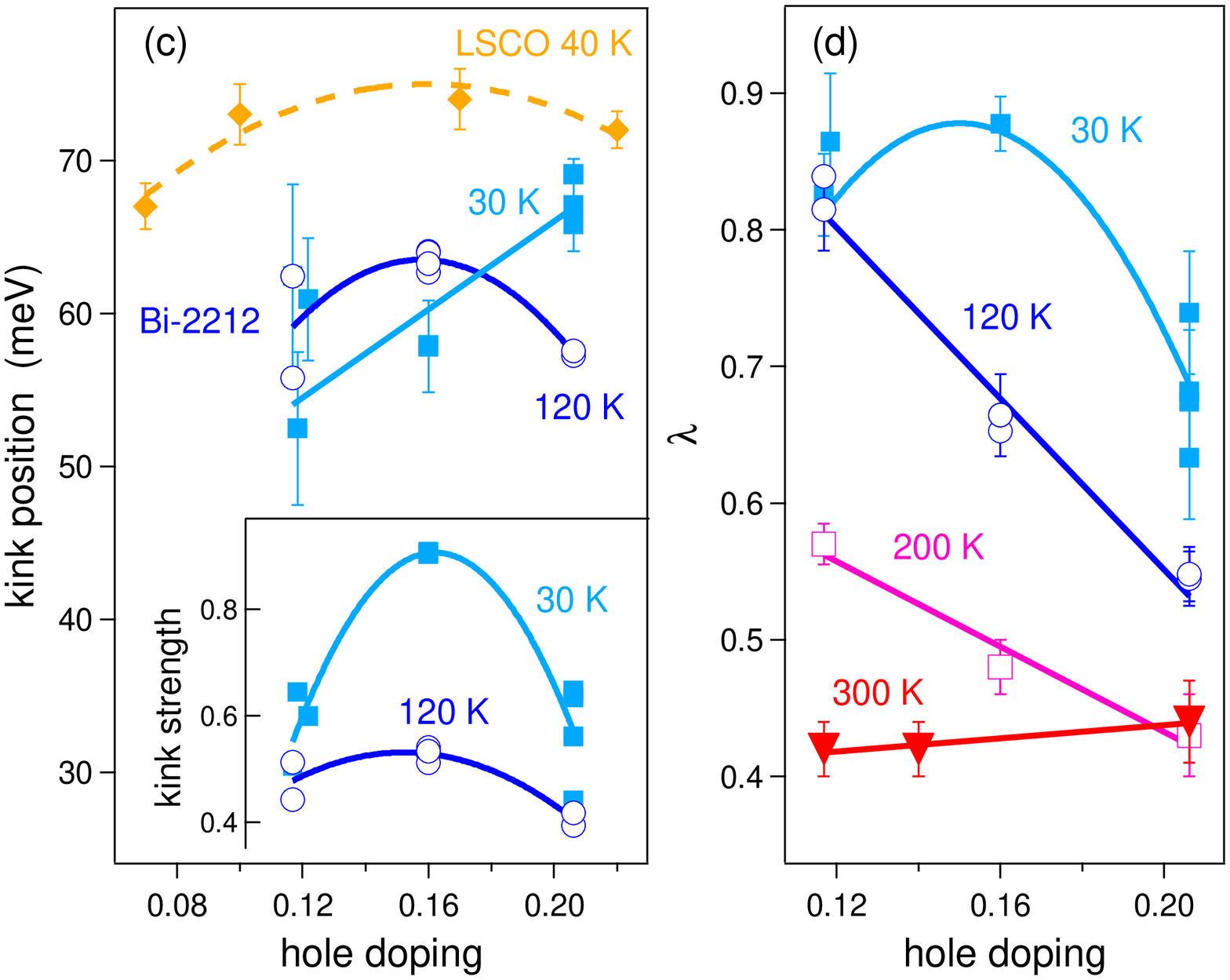}%
\caption{\label{ReS} (a), (b) Self-energy functions derived from experimental data for an overdoped, underdoped, and optimally doped Bi-2212 samples; the inset on (a) shows the data taken $\approx 0.15$ \AA$^{-1}$ away from the node to monitor the gap opening; on (b) the solid curve shows fit of $\Sigma'(30 K)-\Sigma'(120 K)$ to a gaussian. (c), (d) The evolution of the kink parameters with hole concentration and temperature.}
\end{figure*}

Main results are summarized in Fig.~\ref{ReS} (a) and (b) where we present the self-energy for the Bi-2212 samples of three doping levels: underdoped (UD77, $x = 0.11$), overdoped (OD75, $x = 0.20$), and optimally doped (OP92, $x = 0.16$). In panel (a) we plot $\Sigma'(\omega)$ for an underdoped and overdoped samples at different temperatures above $T_c$. The self-energies, almost identical for the room temperatures, become essentially different for lower temperatures (200 K and 120 K). In other words, an increase of $\Sigma'(\omega)$ with lowering the temperature is drastically different for overdoped and underdoped samples. However, in both cases, this increase exhibits a kink close to 60 meV (vertical dashed line). 

The inset illustrates the persistence of the kink feature for the overdoped samples over the superconducting transition. Here we show $\Sigma'(\omega)$ extracted from the spectra taken $\approx 0.15$ \AA$^{-1}$ away from the node to monitor the presence of the superconducting gap, which effects the MDC dispersion at low energy. The position of the kink remains unchanged over the superconducting transition but decreases with further increase of the temperature. Here one can notice some feature around 40 meV that appears with the gap opening. This should be a natural consequence of the gapped density of states and illustrates a rather weak effect of it on the nodal dispersion. 

In panel (b), for  an optimally doped sample, we examine the evolution of both $\Sigma'(\omega)$ and $\Sigma''(\omega)$ comparing the data taken at 30 K and 120 K. The blue and grey shaded areas represent the change in the real and imaginary parts respectively. While the increase in $\Sigma'(\omega)$ with lowering temperature from 300 K to 120 K for the UD and OP samples are dramatic (the room temperature $\Sigma'$ for OP92 is not shown but coincides with the corresponding curves for UD77 and OD75), in the range from 120 K to 30 K only a sharpening of the kink is observed. Note, that while the kink feature on $\Sigma'(\omega)$ at both temperatures stays approximately at the same energy, their difference is peaked at some lower energy ($\approx$ 50 meV, the solid curve shows its fit to a gaussian) that is in agreement to earlier result \cite{JohnsonPRL01}. 

Following the tradition of model independent data treatment we fit the self-energy (in the energy range about 150 meV below $E_F$) to a simple kink-function $-\Sigma'(\omega) = \lambda \omega (|\omega|<\omega_k) + (\lambda_h \omega + C) (|\omega|>\omega_k)$. The fitting parameters---kink position $\omega_k$, kink strength $|\lambda-\lambda_h|$, and coupling strength $\lambda$---are plotted in Fig.~\ref{ReS} (c) and (d) as function of doping and temperature. 

The presented $\Sigma'(\omega)$ dependencies are consistent with the idea of two channels in the scattering process which has been earlier deduced from the qualitative analysis of the scattering rate \cite{KordyukPRL2004}. Due to the results presented here the idea of two channel scattering is not only supported by more careful analysis but also, using the advantage of the KK-consistent procedure \cite{KordyukPRB2005}, can be described quantitatively. In the following interpretation we describe a model consistent with the experimental data.

We distinguish two scattering channels which we mark as \textquotedblleft primary" and \textquotedblleft secondary" ($\Sigma_1$ and $\Sigma_2$). The former is mainly $xT$-independent while the latter exhibits a critical dependence on both temperature and doping. 

\begin{figure}[t]
\includegraphics[width=6.2cm]{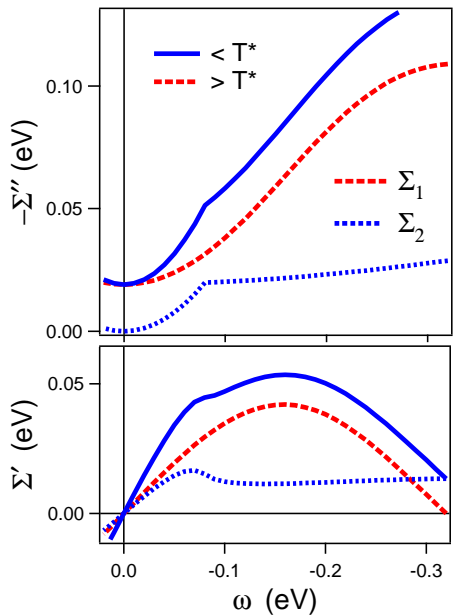}%
\caption{\label{Sketch} Schematic summary of the temperature evolution of the real (bottom) and imaginary (top) parts of the self-energy.}
\end{figure}

The \textit{primary} channel, due to its $xT$-independence and structureless energy dependence, can be naturally associated with the direct electron-electron Coulomb interaction which results in an Auger-like scattering, the process when a hole decays into two holes and one electron \cite{KordyukPRL2004}. In close vicinity to the Fermi level, this process forms the quasiparticles of the Fermi liquid type with $\Sigma'_1(\omega) = -\lambda \omega$ and $\Sigma''_1(\omega) \propto \omega^2 + (\pi T)^2$. In a finite energy range, the self-energy depends on quasiparticle density of states (DOS), following the asymptotic behavior until DOS($\omega$) = const. On a large scale, a confined DOS (with a cut-off at $\sim \omega_c$) leads to a non-monotonic $\Sigma''_1(\omega)$---roughly, it reaches a maximum at $\omega_c$. Also roughly, being KK-related, the real part $\Sigma'_1(\omega)$ reaches its maximum close to an inflexion point of $\Sigma''_1(\omega)$ and at $\omega_c$ goes to zero. This behavior is schematically shown in Fig.~\ref{Sketch} by the red dashed lines. Here we should note that the deduced value $\omega_c \approx$ 0.3 eV is approximately 3 times smaller than the bare band width $\omega_0$ \cite{KordyukPRB2005} that is difficult to explain by simple renormalization of the bare DOS. Possible explanation can be related with the highly non-uniform real DOS due to the van Hove singularities caused by the saddle points and/or with kinematic constraints. The latter seems to be really essential if we recall a very small effect of the gap opening on low energy part of $\Sigma'(\omega)$ (see the inset in Fig.~\ref{ReS}a), so, one can conclude that the effective DOS, which forms the self-energy of nodal quasiparticles, consists of states mainly from the nodal region. Thus, except the temperature dependent offset of the scattering rate, $\Sigma''(0,T)$, the temperature and doping dependence of the primary scattering channel appears only through the effective DOS and remains weak. Roughly, this channel can be described by two $xT$-independent parameters, a coupling strength $\lambda_1 = 0.43\pm0.02$ and scattering cut-off $\omega_{c1} = 0.35\pm0.05$ eV.

The \textit{secondary} channel is essentially different. First, it exhibits very different behavior---it gradually appears only below some temperature, while both this temperature and strength of the channel are progressively increasing with underdoping. Second, the channel is not structureless but reveals some energy scale that implies a certain structure of the interactions involved. It is this structure that forms the \textquotedblleft 70 meV kink", depending on doping (Fig.~\ref{ReS}c) and changing over the superconducting transition (Fig.~\ref{ReS}b).

The shape of $\Sigma''_2(\omega)$ dependence (see Fig.~\ref{Sketch}) indicates a bosonic origin of the secondary scattering channel, a process during which a hole decays into another hole and a bosonic excitation. In a simplest case of constant electronic DOS the coupling to a single bosonic mode would result in a step-like function. In general case, $\Sigma''_2(\omega)$ is a convolution of the bosonic and electronic density of states. Therefore, a careful examination of this channel will not only reveal the origin of relevant bosons but also provide parameters of the bosonic spectrum which can help to understand the nature of the superconducting coupling. Phonons and spin-fluctuations are considered as the most probable candidates for the role of the main scattering bosons \cite{DamascelliRMP03}. In the following we discuss which of them can be consistent with the experimental data, and, since it is not necessary that the secondary channel is formed by bosons of only one type, we split the problem into two parts. (1) We derive the properties of the bosonic spectrum which makes a global contribution into secondary channel. (2) We discuss the structure of the scattering rate, whether it is possible to identify it with bosons of a certain type.

The global contribution to $\Sigma_2(\omega)$ can be evaluated in terms of a coupling strength of this channel $\lambda_2 = \lambda - \lambda_1$. Strong dependence of this parameter on temperature and doping can be seen in Fig.~\ref{ReS} (d): being negligible at room temperature it grows as big as $\lambda_1$ at low temperatures (for UD and OP) and, e.g. at 120 K, it grows from 0.1 to 0.4 when the hole concentration is decreasing from $x =$ 0.21 (OD75) to 0.12 (UD77). At temperatures higher than $T_c$, $\lambda_2$ exhibits monotonic (roughly linear) dependence on doping. We believe that $\lambda_2$ vanishes at $T^*(x)$ that is consistent with the previous qualitative consideration \cite{KordyukPRL2004} although from current results we can only deduce that $T^*(0.21)<$ 200 K $<T^*(0.16)$. The \textquotedblleft 70 meV kink" behaves similarly, or, at least, it vanishes together with $\lambda_2$. This makes us to believe that the kink is an inherent feature of a bosonic spectrum which makes the main contribution to $\Sigma_2(\omega)$.

Common sense suggests that phonons cannot be responsible for such a dramatic doping and temperature dependence of the channel strength. Only softening of some bosonic modes has been reported (e.g., see \cite{McQueeneyPRL2001}) but it seems unlikely that the phononic spectrum can completely disappear with doping or temperature. One can get an additional argument from \textit{purified} $\Sigma''_2(\omega)$ dependence (see Fig.~\ref{Sketch}) which does not saturate up to 300 meV binding energy that is difficult to reconcile with $\sim$ 90 meV confined phononic spectrum \cite{McQueeneyPRL2001}. On the other hand, the shape of $\Sigma''_2(\omega)$ reminds one which is expected for the spin-fluctuation scattering (continuum plus mode) \cite{EschrigPRB2003}, although the energy of the scattering kink in experiment is notably higher: 100 meV \cite{KordyukPRL2004} instead of 70 meV \cite{EschrigPRB2003}. From this \textquotedblleft energy argument", more general assumption that the 100 eV energy scale in scattering, which results is the \textquotedblleft 70 meV kink" in dispersion, is formed by the gaped continuum \cite{ChubukovPRB2004} seems more adequate. Moreover, recently predicted new mode in spin susceptibility \cite{EreminPRL2005} seems to fit very well the presented data. Therefore we may conclude here that the secondary channel in the nodal scattering is mainly caused by an indirect interaction between electrons through the spin-fluctuations.

Remarkable is the dependence of the kink energy on hole concentration, $\omega_{k}(x)$, for $T_c<T<T^*$ (see Fig.~\ref{ReS}c, 120 K), which, like $T_c(x)$, exhibits a maximum at optimal doping level. Similar behavior is observed for the normal state of LSCO measured at 40 K (yellow dashed line in Fig.~\ref{ReS}c). This may signify a certain correlation between the superconducting glue and spin-fluctuation spectrum while the dramatic evolution of $\omega_{k}(x)$ and $\lambda_2(x)$ over superconducting transition indicates a strong correlation of electronic and bosonic spectra. We note that the monotonic $\omega_{k}(x)$ dependence for $T<T_c$ is in agreement with earlier results \cite{GromkoPRB2003}.

Finally, it can happen that the situation is more complex than it is seen. Phonons might make an observable contribution to the kink story, interfering with the spin-fluctuation kink in its dependence on doping and over the superconducting transition. In this case, however, the maximum coupling to the phonons can be estimated as $\lambda_{ph} \sim$ 0.1, that practically rules them out as a glue for the superconducting pairing. If the spin-fluctuations are the main reason for $xT$-dependent scattering in the pseudo-gap state, a contribution of phonons is smaller being probably beyond the accuracy of modern photoemission experiment. Nevertheless it seems highly important to keep going in this direction improving the quality of the experimental data to be able to compare the bosonic spectrum extracted from photoemission to spectra of spin-fluctuations and phonons in order to find out the details of the pairing process.

In conclusion, we distinguish two principal channels of interactions which form the quasiparticle self-energy along the nodal direction of high-$T_c$ cuprates. Both are originated from interaction in electronic subsystem, but while the primary channel is structureless and mainly $xT$-independent and, therefore, can be naturally explained by a direct electron-electron scattering (the Auger process), the secondary channel exhibits a critical dependence on doping and temperature in agreement with spin-fluctuation spectrum and can be explained by an indirect process via the magnetic degree of freedom. While the maximum of the renormalization is related with saturation of the Auger process, the kink feature on the experimental dispersion appears only with underdoping and/or lowering temperature and is caused by an energy scale in the spin-fluctuation spectrum. The evolution of this scale with doping and temperature indicates therefore an intimate relation of the spin-fluctuations with mechanism of high-$T_c$ superconductivity.

The project is part of the Forschergruppe FOR538. The work in Lausanne was supported by the Swiss National Science Foundation and by the MaNEP.

\end{document}